\documentstyle[12pt,epsfig]{article}
\begin{document}
\noindent
\begin{center}
{\Large {\bf Cosmological Constant in Chameleon \\Brans-Dicke Theory }}\\
\vspace{2cm}
 ${\bf Yousef~Bisabr}$\footnote{e-mail:~y-bisabr@srttu.edu.}\\
\vspace{.5cm} {\small{Department of Physics, Shahid Rajaee Teacher
Training University,
Lavizan, Tehran 16788, Iran}}\\
\end{center}
\begin{abstract}
We consider a generalized Brans-Dicke model in which the scalar
field has a self-interacting potential function.  The scalar field is also allowed to couple
non-minimally with the matter part. We assume that it has a chameleon behavior in the sense that it acquires a
density-dependent effective mass.  We consider two different types of matter systems
which couple with the chameleon, dust and vacuum.  In the first case, we find a set of exact solutions
when the potential has an exponential form.  In the second case, we find a power-law
exact solution for the scale factor.  In this case, we will show that the vacuum density
decays during expansion due to coupling with the chameleon.

\end{abstract}

~~~~~~~~~~~~~~~~~~~~~~~~~~~~~~~~~~~~~~~~~~~~~~~~~~~~~~~~~~~~~~~~~~~~~~~~~~~~~~~~~~~~~~~~~~~~~~~~~~~~~
\section{Introduction}
One of the approaches to address recent problems in standard
cosmology is to
attribute these problems to some modifications of general
relativity. Such modified gravity models can be obtained in
different ways. For instance, one can replace the Ricci scalar in
the Einstein-Hilbert action by some functions $f(R)$ (for a review
see, e.g., \cite{1} \cite{cf} and references therein) the so-called $f(R)$
gravity models.  Another approach is scalar-tensor gravity \cite{2}
which associates a scalar partner to the metric tensor for
describing geometry of spacetime . The prototype of the latter is
Brans-Dicke (BD) theory \cite{BD} which its original motivation was
the search for a theory containing Machs principle. As the simplest
and best-studied generalization of general relativity, it is natural
to think about the BD scalar field as a possible candidate for
producing modifications without invoking auxiliary fields or exotic
matter systems.  It is shown that this theory can actually produce a
non-decelerating expansion for low negative values of the BD
parameter \cite{b}. Unfortunately, this conflicts with the lower
bound imposed on this parameter by solar system experiments
\cite{w}.\\
 There has recently been a tendency in the
literature  \cite{4} \cite{5} \cite{6} \cite{6a} to consider the BD scalar
field as a chameleon \cite{k}. In these models, the scalar field can be heavy
enough in the environment of the laboratory tests so that the local
gravity constraints suppressed. Meanwhile, it can be light enough in
the low-density cosmological environment to be considered as a
candidate for dark energy. These different behaviors in small and
large scales depend crucially on the shapes of the potential and the
coupling functions since both functions contribute to the effective
mass or compton wavelength of the scalar field.  It is shown that
these requirements can be fulfilled by choosing an appropriate
potential function for the scalar field
\cite{k}. We have already shown that such a chameleon BD theory can be potentially
consistent with observations for large and positive BD parameter \cite{6}.  \\
In the present work, we will focus on the late-time behavior of the Universe and the cosmological constant
problem.
This work is organized as follows : In section 2, we present the
basic equations and definitions of the model. We
apply these equations in a cosmological setting. The matter system is not conserved due to its interaction with chameleon.
We solve the (non-) conservation equation.  The solution indicates a modification of standard evolution of matter density in terms of a parameter
which quantifies the energy transfer.
We also solve the field equations for exponential potentials in a matter-dominated case.
 In section 3, we introduce a decaying mechanism for a large cosmological constant based on the non-minimal coupling of
vacuum energy density.  In section 4, we draw our conclusions.

~~~~~~~~~~~~~~~~~~~~~~~~~~~~~~~~~~~~~~~~~~~~~~~~~~~~~~~~~~~~~~~~~~~~~~~~~~~~~~~~~~~~~~~~~~~~~~~~~~~~~~~~~~~~~~~~~~~~~~~~~~~~~~~~
\section{The Cosmological Setting}
Let us begin with the following action
\begin{equation}
S=\frac{1}{16\pi}\int d^4x \sqrt{-\bar{g}} \{\phi
\bar{R}-\frac{\omega_{BD}}{\phi}\bar{g}^{\mu\nu}\bar{\nabla}_{\mu}\phi
\bar{\nabla}_{\nu}\phi -V(\phi)+16\pi f(\phi)L_m\}
\label{1}\end{equation} where $\bar{R}$ is the Ricci scalar, $\phi$
is the BD scalar field which we also take it to be a chameleon
field, $\omega_{BD}$ is the BD parameter, $V(\phi)$ and $f(\phi)$ are some analytic functions.  The
matter Lagrangian density, which is denoted by $L_m$, is coupled
with $\phi$ by the function $f(\phi)$. If $f(\phi)=1$, we return to
BD action with
a potential function $V(\phi) ~\cite{2}$.  \\
A conformal transformation
\begin{equation}
\bar{g}_{\mu\nu}\rightarrow g_{\mu\nu}=\Omega^2 \bar{g}_{\mu\nu}
\label{a2}\end{equation} with $\Omega=\sqrt{G \phi}$ brings the
above action into the Einstein frame \cite{1} \cite{far1}.  If we
also redefine the scalar field
\begin{equation}
\varphi(\phi)=\sqrt{\frac{2\omega_{BD}+3}{16\pi G}}\ln
(\frac{\phi}{\phi_0}) \label{a3}\end{equation} with $\phi_0\sim
G^{-1}$, $\phi>0$ and $\omega_{BD}>-\frac{3}{2}$ then the kinetic term of
the scalar field takes a canonical form.  In terms of the new
variables ($g_{\mu\nu}$, $\varphi$) the action (\ref{1}) takes then
the form
\begin{equation}
S_{EF}= \int d^{4}x \sqrt{-g} \{\frac{R}{16\pi G}
-\frac{1}{2}g^{\mu\nu}\nabla_{\mu}\varphi
\nabla_{\nu}\varphi-U(\varphi)+
\exp(-\frac{\sigma\varphi}{M_p})~f(\varphi) L_{m}\}
\label{a5}\end{equation} where $\sigma=8\sqrt{\frac{\pi
}{2\omega+3}}$ and $\nabla_{\mu}$ is the covariant derivative of the
rescaled metric $g_{\mu\nu}$.  The Einstein frame potential is
\begin{equation}
U(\varphi)= V(\phi(\varphi))~\exp(-\sigma\varphi/M_{p})
\label{a6}\end{equation} Variation of the action (\ref{a5}) with
respect to the metric $g_{\mu\nu}$ and $\varphi$ gives,
respectively,
\begin{equation}
G_{\mu\nu}=M_p^{-2} (h(\varphi)T^{m}_{\mu\nu}+T^{\varphi}_{\mu\nu})
\label{2}\end{equation}
\begin{equation}
\Box \varphi-U'(\varphi)=-h'(\varphi) L_m \label{3}\end{equation}
where
\begin{equation}
T^{\varphi}_{\mu\nu}=(\nabla_{\mu}\varphi
\nabla_{\nu}\varphi-\frac{1}{2} g_{\mu\nu}\nabla_{\alpha}\varphi
\nabla^{\alpha}\varphi) -U(\varphi)g_{\mu\nu}
\label{4}\end{equation}
\begin{equation}
T^m_{\mu\nu}=\frac{-2}{\sqrt{-g}}\frac{\delta (\sqrt{-g}L_m)}{\delta
g^{\mu\nu}} \label{5}\end{equation} and
$h(\varphi)=e^{-\sigma\varphi/M_{p}}f(\varphi)$. From now on, prime indicates
differentiation with respect to $\varphi$. One can write (\ref{3})
in the form
\begin{equation}
\nabla^\mu T^{\varphi}_{\mu\nu}=-\nabla_{\nu}h(\varphi)L_m
\label{6}\end{equation} or, equivalently,
\begin{equation}
\dot{\rho}_{\varphi}+3\frac{\dot{a}}{a}(\omega_{\varphi}+1)\rho_{\varphi}=h'(\varphi)\dot{\varphi}L_m
\label{3a}\end{equation}
 where $\omega_{\varphi}=p_{\varphi}/\rho_{\varphi}$, $\rho_{\varphi}=\frac{1}{2}\dot{\varphi}^2+U(\varphi)$ and
$p_{\varphi}=\frac{1}{2}\dot{\varphi}^2-U(\varphi)$.  As a result of
the explicit coupling of matter system with the scalar field,
covariant derivative of matter stress-tensor $T^m_{\mu\nu}$ does not
vanish. This can be seen by applying the Bianchi identities
$\nabla^{\mu}G_{\mu\nu}=0$ to (\ref{2}), which leads to
\begin{equation}
\nabla^{\mu}T^m_{\mu\nu}=(L_m-T^m)\nabla_{\nu}\ln h(\varphi)
\label{6}\end{equation}
where $T^m=g^{\mu\nu}T^m_{\mu\nu}$.  It is clear from (\ref{6}) that details of
the energy exchange between matter and $\varphi$ depends on the
explicit form of the matter Lagrangian density $L_m$.  Here we
consider a perfect fluid energy-momentum tensor as a matter system
\begin{equation}
T^m_{\mu\nu}=(\rho_m+p_m)u_{\mu}u_{\nu}+p_mg_{\mu\nu}
\label{b1}\end{equation}
where $\rho_m$ and $p_m$ are energy density and pressure, respectively. The four-velocity of the fluid is denoted by $u_{\mu}$.\\
There are different choices for the perfect fluid Lagrangian density
which all of them leads to the same energy-momentum tensor and field
equations in the context of general relativity \cite{sh} \cite{haw}.
The two Lagrangian densities that have been widely used in the
literature are $L_m=p_m$ and $L_m=-\rho_m$ \cite {o} \cite{ber}
\cite{s}.  For a perfect fluid that does not couple explicitly to
the curvature (i.e., for $f(\varphi) = 1$), the two Lagrangian
densities $L_m =p_m$ and $L_m=-\rho_m$ are perfectly equivalent, as
discussed in \cite{ber} \cite{s}. However, in the model presented
here the expression of $L_m$ enters explicitly the field equations
and all results strongly depend on the choice of $L_m$.  In fact, it
is shown that there is a strong debate about equivalency of
different expressions attributed to the Lagrangian density of a
coupled perfect fluid \cite{far}. Here we take $L_m=p_m$ for the
lagrangian density.\\ We apply the field equations (\ref{2}) and
(\ref{3}) to a spatially flat Friedmann-Robertson-Walker spacetime
\begin{equation}
ds^2=-dt^2+a^2(t)(dx^2+dy^2+dz^2) \label{a10}\end{equation} with
$a(t)$ being the scale factor.  This gives
\begin{equation}
3H^2=M_p^{-2}(h(\varphi)\rho_{m}+\rho_{\varphi})
\label{a11}\end{equation}
\begin{equation}
2\dot{H}+3H^2=-M_p^{-2}~(h(\varphi)p_{m}+p_{\varphi})
\label{a12}\end{equation}
\begin{equation}
\ddot{\varphi}+3H\dot{\varphi}+\frac{dU(\varphi)}{d\varphi}=p_{m}~h'(\varphi)
\label{a13}\end{equation}
where $H=\frac{\dot{a}}{a}$.   Moreover, the conservation equations
become
\begin{equation}
\dot{\rho}_{\varphi}+3H(\omega_{\varphi}+1)\rho_{\varphi}=h'(\varphi)\dot{\varphi}~\rho_m
\label{a15}\end{equation}
\begin{equation}
\dot{\rho}_{m}+3H(\omega_m+1)\rho_m=-(1-2\omega_m)\dot{\varphi}\frac{h'}{h}
\rho_m \label{a14}\end{equation}
 The latter can be solved which gives the following solution
\begin{equation}
\rho_m=\rho_{m0} a^{-3(\omega_m+1)}e^{-\frac{(1-2\omega_m)}{M_p}\int
\beta(\varphi)d\varphi}\label{1a16}\end{equation} where $\frac{h'(\varphi)}{h(\varphi)}\equiv \frac{\beta(\varphi)}{M_p}$ and
$\rho_{m0}$ is an integration constant. It can also be written as
\cite{6} \cite{bis}
\begin{equation}
\rho_m=\rho_{m0} a^{-3(\omega_m+1)+\varepsilon}
\label{a166}\end{equation} where we have defined
\begin{equation}
\varepsilon\equiv \frac{(2\omega_m-1)}{M_p}\frac{\int \beta d\varphi}{\ln
a}\label{a16}\end{equation}
 This solution indicates that the
evolution of energy density is modified due to interaction of
$\varphi$ with matter.  It states that when $\varepsilon<0$, matter
is created and energy is injecting into the matter so that the
latter will dilute more slowly compared to its standard evolution
$\rho_m\propto a^{-3(\omega_m+1)}$. Similarly, when $\varepsilon>0$
the reverse is true, namely that matter is annihilated and the
direction of energy transfer is outside of the matter system so that
the rate of dilution is faster than the standard one.\\
Even though $\beta$ and $\varepsilon$ in (\ref{a16}) are generally evolving functions, we consider
the case that they can
be regarded as constant parameters.  The constancy of $\beta$ means that we fix the arbitrary coupling function
$h(\varphi)$ to have an exponential form $h(\varphi)=e^{\beta\frac{\varphi}{M_p}}$.  On the other hand, when $\varepsilon$ is taken to be a constant parameter
the energy transfer between the chameleon and the matter system is constant.  This choice as a first approximation is more restrictive
but greatly simplifies
the mathematics and does not affect the basics of our arguments. In these cases, (\ref{a16}) reduces to
\begin{equation}
\frac{\varphi}{M_p}=\gamma \ln a
\label{c2}\end{equation}
with $\gamma$ being a constant defined by the relation $\epsilon=\beta \gamma(2\omega_m-1)$.
In order to find $a(t)$ and $\varphi(t)$ one should first fix the
potential as an input and solve then (\ref{a11}), (\ref{a12}) and (\ref{a13}).  Here we would like to follow a different
strategy used in \cite{sol}.   For a given
$\varphi(t)$, we look for the potential and the function $a(t)$ that
satisfy the field equations. In this way, one can reduce the second order differential equation of $\varphi$ to a
first order one for finding the functional form of the potential. To do this, we first combine
(\ref{a11}) and (\ref{a12}) to obtain
\begin{equation}
\dot{H}+3H^2=\frac{1}{2}M_p^{-2}~[h(\varphi)(\rho_m-p_m)+(\rho_{\varphi}-p_{\varphi})]
\label{aa12}\end{equation}
We then put the solution (\ref{c2}) into (\ref{a13}) which leads to
\begin{equation}
\dot{H}+3H^2=\frac{p_m}{\gamma M_p}h'(\varphi)-\frac{1}{\gamma M_p}\frac{dU}{d\varphi}
\label{aaa12}\end{equation}
Combining the latter with (\ref{aa12}) gives a consistency relation
\begin{equation}
\frac{dU}{d\varphi}+\frac{\gamma}{M_p}U=\alpha~ h(\varphi)\rho_m
\label{aa13}\end{equation}
The parameter $\alpha$ is defined by $\alpha\equiv [\frac{\beta}{M_p}\omega_m+\frac{\gamma}{2M_p}(\omega_m-1)]$.
This is a first order differential equation which leads to the following solution
\begin{equation}
U(\varphi)=C_1 e^{-\frac{\gamma}{M_p}\varphi}+ \frac{\alpha \rho_{0m}M_p}{\delta+\gamma}~ e^{\frac{\delta}{M_p} \varphi}
\label{aaa12}\end{equation}
where
$
\delta=\beta(2\omega_m-1)-\frac{3}{\gamma}(\omega_m+1)
$
and $C_1$ is an integration constant.  Putting this potential for $\omega_m=0$ into the Friedman equation (\ref{a11}), gives
\begin{equation}
H^2=\frac{1}{M_p^2(3-\gamma^2/2)}\{ \rho_{0m} a^{-3}-\frac{\gamma \rho_{0 m}}{2(\gamma+\delta)} a^{-3-\beta\gamma}
+C_1 a^{-\gamma^2}\}
\end{equation}
The solution for $C_1=0$ and up to an integration constant is
$$
\frac{{2a^{\beta\gamma+2}}\{(1-\frac{\gamma+2\delta}{\gamma}a^{\beta\gamma-1})[(1-\frac{\gamma}{2(\gamma+\delta)})a^{-1}- \frac{\gamma}{2(\gamma+\delta)}
a^{-\beta\gamma}]\}^{\frac{1}{2}}}
{(2+\beta\gamma)[(1-\frac{\gamma}{2(\gamma+\delta)}) a^{\beta\gamma}-\frac{\gamma}{2(\gamma+\delta)} a]}
2F1[\frac{1}{2}, \frac{2+\beta\gamma}{2(\beta\gamma-1)}, \frac{3\beta\gamma}{2(\beta\gamma-1)},
$$
\begin{equation}
 \frac{\gamma+2\delta}{\gamma}a^{\beta\gamma-1}]
=t~~~~~~~~~~~~~~~~~~~~~~~~~~~~~~~~~~~~~~~~~~~~~~~~~~~~~~~~~~~~~~~~~~~~~~~~~~
\end{equation}
where $2F1[a,b,c,x]$ is the hypergeometric function $_2F_1[a,b;c;x]$.
\section{Cosmological constant}
The cosmological constant problem concerns with a large discrepancy between observations
and theoretical predictions on vacuum energy density \cite{c1}. Most of the attempts
trying to resolve this problem are based on the belief that cosmological constant may not have such an extremely small
value suggested by observations at all times and there should exist a dynamical mechanism working during evolution of
the Universe which provides a cancelation of vacuum energy density at late times \cite{c2}.
There is a tendency in Literature to attribute such a reduction to some interactions
with other energy components in the Universe \cite{c3}.
Following this strategy, we intend to relate this reduction to gravitational coupling of vacuum
energy. In the context of chameleon BD theory, we consider coupling of vacuum with the chameleon field.
We show that this interaction can lead to decaying of vacuum energy during expansion of the Universe.  To do this, we first combine (\ref{a11}) and (\ref{a12}) to write
\begin{equation}
\dot{H}=-\frac{1}{2}M_p^{-2}~[h(\varphi)(\rho_m+p_m)+(\rho_{\varphi}+p_{\varphi})]
\label{aaa12}\end{equation}
By taking the matter system as vacuum energy with $p_m\rightarrow p_{\Lambda}$, $\rho_m \rightarrow \rho_{\Lambda}$ and the equation of state $p_{\Lambda}/\rho_{\Lambda}=-1$, (\ref{aaa12}) gives
\begin{equation}
\dot{H}=-\frac{\dot{\varphi}^2}{2M_p^2}
\end{equation}
This together with (\ref{c2}) leads to
\begin{equation}
\dot{H}=-\frac{\gamma^2}{2}H^2
\end{equation}
The solution is $H=\frac{2}{\gamma^2}t^{-1}$ or $a(t)=t^{\frac{2}{\gamma^2}}$.  For this solution, we have
\begin{equation}
h(\varphi)\rho_m\rightarrow h(\varphi)\rho_{\Lambda}= h(\varphi)\rho_{i\Lambda} =a^{\beta\gamma}\rho_{i\Lambda}
\label{ab}\end{equation}
where $\rho_{i\Lambda}$ is the energy density corresponding to a large cosmological constant injected into the Universe at early times.  For $\beta\gamma<0$, this large energy density decays during expansion of the Universe. We require that
$a^{\beta\gamma}\rho_{i\Lambda}$ reduces to $\rho_{0\Lambda}$ at late times, namely that $a^{\beta\gamma}\rho_{i\Lambda}\rightarrow \rho_{0\Lambda}$, with $\rho_{0\Lambda}$ being the present vacuum energy density.  Thus, normalizing the present value
of the scale factor to unity, the relation (\ref{ab}) is equivalent to
\begin{equation}
 a^{\beta\gamma}(t_i)\rho_{i\Lambda}= a^{\beta\gamma}(t_0)\rho_{0\Lambda}
\end{equation}
where $t_i$ is some initial time such as the Planck time and $t_0$ is the age of the Universe. This ensures that $\rho_{i\Lambda}$ decays during expansion to a sufficiently small value
at late-time consistent with observations. Using the power-law solution $a(t)=t^{\frac{2}{\gamma^2}}$ for the
scale factor, we obtain
\begin{equation}
\frac{\rho_{i\Lambda}}{\rho_{0\Lambda}}=(\frac{a(t_0)}{a(t_i)})^{\beta\gamma}=(\frac{t_0}{t_i})^{\frac{2\beta}{\gamma}}
\end{equation}
$$
~~~~~~~~~~~~=(\frac{10^{17}sec}{10^{-43}sec})^{\frac{2\beta}{\gamma}}=(10^{60})^{\frac{2\beta}{\gamma}}
$$
Combining the latter with $\frac{\rho_{i\Lambda}}{\rho_{0\Lambda}}\sim 10^{120}$, constrains the ratio $\frac{\beta}{\gamma}$ to be of order of unity.  The decay of $a^{\beta\gamma}\rho_{i\Lambda}$ is depicted in fig.1 for different values of the parameters $\beta$ and $\gamma$.\\
This decaying mechanism is also consistent with (\ref{a166}) characterizing energy transfer between  matter perfect fluid and chameleon.  This can be easily shown by using (\ref{a16}) and writing (\ref{a166}) for vacuum $a^{3\beta\gamma}\rho_{\Lambda}=\rho_{0\Lambda}$.  For $\beta\gamma<0$, this relation indicates transferring energy from vacuum to chameleon during expansion.
\begin{figure}[ht]
\begin{center}
\includegraphics[width=0.6\linewidth]{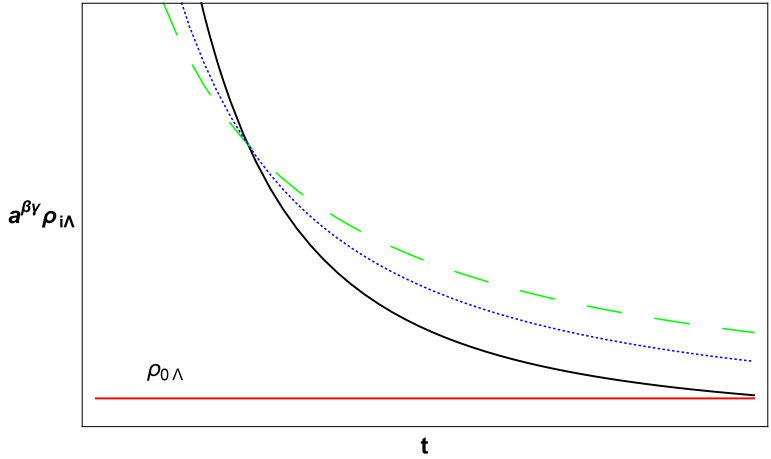}
\caption{The plot of $a^{\beta\gamma}\rho_{i\Lambda}$ for $\beta=-1$, $\gamma=1$ (solid), $\beta=-1/2$, $\gamma=1$ (dashed) and
$\beta=-1/3$, $\gamma=1/2$ (dotted).}
\end{center}
\end{figure}

~~~~~~~~~~~~~~~~~~~~~~~~~~~~~~~~~~~~~~~~~~~~~~~~~~~~~~~~~~~~~~~~~~~~~~~~~~~~~~~~~~~~~~~~~~~~~~~~~~~~~~~~~~~~~~~~~~~~~~~~~~~~~~~

\section{Conclusion}
In this work we have investigated some features of BD theory with a
self-interacting potential.  Gravitational coupling of matter is usually constrained to be minimal
in order to keep a gravitational model in accord with weak equivalence principle.  Here this constraint is relaxed and
BD scalar field is allowed to couple
non-minimally with the matter part. Non-minimal gravitational coupling does not necessarily
mean violation of weak equivalence principle and there is still a possibility that the effective mass
of the scalar field be scale dependent.  The scalar field is taken to be a chameleon
field in order to pass local gravity experiments at late times.\\ In this chameleon BD theory the matter system is not conserved due to interaction with the
chameleon.  We solved the matter conservation equation and formulated the solution
as $\rho_m\propto a^{-3(\omega_m+1)+\varepsilon}$.  This explicitly indicates the energy transfer between the two components. Throughout our analysis we assumed an exponential form for the coupling function
($\beta$ is a constant parameter) and a constant rate of energy transfer ($\varepsilon$ is a constant parameter).  The
main results of the analysis are the following:\\
1)We found a set of exact solutions for the field equations for exponential potentials. \\
2) The non-minimal coupling of chameleon with the matter part motivated us to construct a mechanism for decaying a large
cosmological constant.  Taking the matter system as a perfect fluid with equation of state $\omega_{\Lambda}=-1$, we found a power-law
form for evolution of the scale factor $a(t)=t^{\frac{2}{\gamma^2}}$ implying accelerating expansion for $\gamma^2<2$.  We have shown that
interaction of vacuum with the chameleon can actually make the former reduce during expansion of the Universe.  The late-time value of vacuum density is sufficiently small according to observations if $\frac{\beta}{\gamma}\sim 1$.

~~~~~~~~~~~~~~~~~~~~~~~~~~~~~~~~~~~~~~~~~~~~~~~~~~~~~~~~~~~~~~~~~~~~~~~~~~~~~~~~~~~~~~~~~~~~~~~

\end{document}